\documentclass{JHEP3}

\usepackage{epsfig}
\def \be {\begin{equation}}
\def \ee {\end{equation}}
\def \bea {\begin{eqnarray}}
\def \eea {\end{eqnarray}}
\def \nn {\nonumber}

\def \a {\alpha}
\def \b {\beta}
\def \g {\gamma}
\def \G {\Gamma}
\def \d {\delta}

\def \m {\mu}
\def \n {\nu}
\def \k {\kappa}

\def \s {\sigma}
\def \r {\rho}
\def \o {\omega}
\def \O {\Omega}
\def \th {\theta}
\def \Th {\Theta}

\def \t {\tau}
\def \dag {\dagger}
\def \p {\partial}

\def\bd{\begin{document}}
\def\ed{\end{document}}
\def\nn{\nonumber}
\def\bea{\begin{eqnarray}}
\def\eea{\end{eqnarray}}
\let\bm=\bibitem
\let\la=\label

\def\N{{\cal N}}
\def\sst{\scriptscriptstyle}
\def\thetabar{\bar\theta}
\def\Tr{{\rm Tr}}
\def\one{\mbox{1 \kern-.59em {\rm l}}}

\def\a{\alpha}      \def\da{{\dot\alpha}}
\def\b{\beta}       \def\db{{\dot\beta}}
\def\c{\gamma}  \def\C{\Gamma}  \def\cdt{\dot\gamma}
\def\d{\delta}  \def\D{\Delta}  \def\ddt{\dot\delta}
\def\e{\epsilon}        \def\vare{\varepsilon}
\def\f{\phi}    \def\F{\Phi}    \def\vvf{\f}
\def\h{\eta}
\def\k{\kappa}
\def\l{\lambda} \def\L{\Lambda}
\def\m{\mu} \def\n{\nu}
\def\o{\omega}
\def\P{\Pi}
\def\r{\rho}
\def\s{\sigma}  \def\S{\Sigma}
\def\t{\tau}
\def\th{\theta} \def\Th{\Theta} \def\vth{\vartheta}
\def\X{\Xeta}
\def\z{\zeta}
\def\w{\wedge}
\def\u{\underline}
\def\hs{\hspace}

\def\cA{{\cal A}} \def\cB{{\cal B}} \def\cC{{\cal C}}
\def\cD{{\cal D}} \def\cE{{\cal E}} \def\cF{{\cal F}}
\def\cG{{\cal G}} \def\cH{{\cal H}} \def\cI{{\cal I}}
\def\cJ{{\cal J}} \def\cK{{\cal K}} \def\cL{{\cal L}}
\def\cM{{\cal M}} \def\cN{{\cal N}} \def\cO{{\cal O}}
\def\cP{{\cal P}} \def\cQ{{\cal Q}} \def\cR{{\cal R}}
\def\cS{{\cal S}} \def\cT{{\cal T}} \def\cU{{\cal U}}
\def\cV{{\cal V}} \def\cW{{\cal W}} \def\cX{{\cal X}}
\def\cY{{\cal Y}} \def\cZ{{\cal Z}}

\def\ua{\underline{\alpha}} \def\ubb{\underline{\beta}}
\def\ug{\underline{\gamma}}
\def\ub{\underline{\phantom{\alpha}}\!\!\!\beta}
\def\uc{\underline{\phantom{\alpha}}\!\!\!\gamma}
\def\um{\underline{\mu}} \def\un{\underline{\nu}}
\def\ud{\underline\delta}
\def\ue{\underline\epsilon}
\def\una{\underline a}\def\unA{\underline A}
\def\unb{\underline b}\def\unB{\underline B}
\def\unc{\underline c}\def\unC{\underline C}
\def\und{\underline d}\def\unD{\underline D}
\def\une{\underline e}\def\unE{\underline E}
\def\unf{\underline{\phantom{e}}\!\!\!\! f}\def\unF{\underline F}
\def\unm{\underline m}\def\unM{\underline M}
\def\unn{\underline n}\def\unN{\underline N}
\def\unp{\underline{\phantom{a}}\!\!\! p}\def\unP{\underline P}
\def\unq{\underline{\phantom{a}}\!\!\! q}
\def\unQ{\underline{\phantom{A}}\!\!\!\! Q}
\def\unH{\underline{H}}

\def\As {{A \hspace{-6.4pt} \slash}\;}
\def\bs {{b \hspace{-6.4pt} \slash}\;}
\def\Ds {{D \hspace{-6.4pt} \slash}\;}
\def\ds {{\del \hspace{-6.4pt} \slash}\;}
\def\ss {{\s \hspace{-6.4pt} \slash}\;}
\def\ks {{ k \hspace{-6.4pt} \slash}\;}
\def\ps {{p \hspace{-6.4pt} \slash}\;}
\def\pas {{{p_1} \hspace{-6.4pt} \slash}\;}
\def\pbs {{{p_2} \hspace{-6.4pt} \slash}\;}

\def\Fh{\hat{F}}
\def\Vh{\hat{V}}
\def\Xh{\hat{X}}
\def\ah{\hat{a}}
\def\xh{\hat{x}}
\def\yh{\hat{y}}
\def\ph{\hat{p}}
\def\xih{\hat{\xi}}

\def\psit{\tilde{\psi}}
\def\Psit{\tilde{\Psi}}
\def\tht{\tilde{\th}}

\def\At{\tilde{A}}
\def\Qt{\tilde{Q}}
\def\Rt{\tilde{R}}
\def\Nt{\tilde{N}}

\def\at{\tilde{a}}
\def\st{\tilde{s}}
\def\ft{\tilde{f}}
\def\pt{\tilde{p}}
\def\qt{\tilde{q}}
\def\vt{\tilde{v}}
\def\nt{\tilde{n}}

\def\delb{\bar{\partial}}
\def\bz{\bar{z}}
\def\bD{\bar{D}}
\def\bB{\bar{B}}

\def\bk{{\bf k}}
\def\bl{{\bf l}}
\def\bp{{\bf p}}
\def\bq{{\bf q}}
\def\br{{\bf r}}
\def\bx{{\bf x}}
\def\by{{\bf y}}
\def\bR{{\bf R}}
\def\bV{{\bf V}}

\def\d{\delta}\def\D{\Delta}\def\ddt{\dot\delta}

\def\p{\partial} \def\del{\partial}
\def\xx{\times}
\def\uno{\mbox{1 \kern-.59em {\rm l}}}

\def\trp{^{\top}}
\def\inv{^{-1}}
\def\dag{{^{\dagger}}}
\def\pr{\prime}

\def\rar{\rightarrow}
\def\lar{\leftarrow}
\def\lrar{\leftrightarrow}

\def\r06{\epsilon^6}
\def\ta{\tilde a}

\title{Wilson-Polyakov surfaces and M-theory branes}

\author{Bin Chen\\
Department of Physics,\\
Peking University,\\
Beijing 100871, P.R. China\\
 \email{bchen01@pku.edu.cn}}

\author{Jun-Bao Wu\\International School for Advanced Studies (SISSA), and INFN,\\
via Beirut 2-4, I-34014 Trieste, Italy\\
 \email{wujunbao@sissa.it}}

\date{\today}

\abstract{In this paper, we study the M-brane description of the
Wilson-Polyakov surfaces in six-dimensional $(2, 0)$ field theory
at finite temperature. We investigate the membrane solution dual
to a straight Wilons-Polyakov surface and compute the interaction
potential between two parallel straight strings by using AdS/CFT
correspondence. Furthermore we discuss the M$5$-brane solutions
dual to various Wilson-Polyakov surfaces. Finally we obtain a
universal result about M5-brane solutions in generic backgrounds.}

\preprint{\  SISSA-07/2008/EP}

\newpage
\begin{document}

\section{Introduction}

The six-dimensional (2,0) superconformal field theory (SCFT) is
mysterious, but also very interesting. From quantum field theory
point of view, it has been known for many years that there exist no
non-trivial unitary superconformal field theory in the spacetime
higher than four dimension, with the six-dimensional one being the
only exception. The existence of a six-dimensional superconformal
field theory was first pointed out in Nahm's beautiful
paper\cite{Nahm:1977tg}. The result was obtained by studying the
representation of superconformal algebra in various
dimensions\cite{Nahm:1977tg}. The same issue was readdressed in
\cite{Seiberg} from the point of view of scaling invariance.  The
superconformal field theory in six-dimensional spacetime has $(2,
0)$ supersymmetries and its field content is just of a tensor
multiplet which includes a two-form $B_{\mu\nu}$ with self-dual
field strength, $4$ fermions and $5$ scalars. Because of the
self-dual two form field, there is no Lagrangian formulation of this
quite mysterious theory\cite{Witten:2007ct}, although this theory is
still a local interacting field theory \cite{Witten96}. After
compactified on a two-torus, the six-dimensional field theory gives
us the four-dimensional ${\cal N}=4$ super Yang-Mills theory at the
low energy limit\cite{Witten:1995zh, Witten:2007ct}. This fact can
be used to study the properties of this very notable four
dimensional superconformal field theory, such as S-duality. In
string theory, this six-dimensional SCFT appears in several
contexts. It appears when we consider IIB string theory on a K3
surface with A-D-E type singularity\cite{Witten:1995zh} and also
appears as the low energy effective field theory of
M$5$-branes\cite{Strominger:1995ac, Witten95}.  In the latter case,
if we consider $N$ M$5$-branes on top of each other, the low energy
effective field theory is the six-dimensional $A_{N-1}, (2, 0)$
SCFT. The five scalars in this field theory describe the
fluctuations of the M$5$-branes in the transverse directions. Since
there is no Lagrangian formulation of the theory, people tried to
study this theory from other angles. A DLCQ matrix model description
of the six-dimensional superconformal field theory has been
suggested \cite{Seiberg97, aharony}, during the development of the
BFSS matrix theory \cite{Banks:1996vh}.

There is another description of M$5$-branes \cite{Gueven:1992hh}:
they can be described as solutions of eleven dimensional
supergravity, which is the low energy effective theory of
M-theory.
Taking the near horizon limit of the supergravity solution gives
us $AdS_7\times S^4$ background with 4-form flux filling in $S^4$.
This led Maldacena to propose the conjecture that the M-theory on
$AdS_7\times S^4$ is dual to the large N limit of the six
dimensional superconformal field theory \cite{Mal97}. This $AdS_7/CFT_6$
correspondence is a cousin of much more well-known $AdS_5/CFT_4$
correspondence proposed in the same paper.
The weak version of the above $AdS_7/CFT_6$ correspondence states
that the large $N$ limit of six-dimensional $(2,0)$ SCFT is dual
to the eleven-dimensional supergravity on $AdS_7 \times S^4$. This
correspondence gives us a new way to study the six-dimensional
theory. The chiral primary operators of the SCFT and the
corresponding supergravity modes were studied in \cite{AdS7CFT}.
Some correlation functions of these local operators were computed
in \cite{Corrado99} from $AdS$ supergravity. These operators were
also studied by using M5-brane action in
\cite{Nurmagambetov:2001ab}.


Non-local operators play important roles in AdS/CFT correspondence.
In the \linebreak $AdS_5/CFT_4$ correspondence, the Wilson loops in
fundamental representation or low dimensional representation can be
described using the fundamental strings \cite{Rey:1998ik, Mal98}.
However, it turns out that the better descriptions of the BPS Wilson
loops in higher dimensional representations are in terms of D-branes
in $AdS_5\times S^5$ \cite{Drukker, Yamaguchi:2006D5, Gomis:2006sb,
Gomis:2006im} due to dielectric effect\cite{Myers:1999}: D3-branes
if the Wilson loops being in symmetric representations, or D5-branes
if being in antisymmetric representations. The D-brane description
of Wilson-'t Hooft operators was discussed in \cite{ChenHe}. It is
remarkable that the D-branes description of  the BPS Wilson loops
encodes the information of string interactions.

A quite similar picture appears also in the $AdS_7/CFT_6$
correspondence. Due to existence of the self-dual 2-form potential,
there are strings in the field theory minimally coupled to the
2-form potential. This allows us to define a two-dimensional
non-local operator called Wilson surface. It can be formally defined
as \cite{Ganor}:
 \be
 W_0(\Sigma)={\rm Tr}\left(\exp i\int_\Sigma B^+\right),\label{ws}
 \ee
where $\Sigma$ is a surface in the six-dimensional spacetime. The
Wilson surface in low dimensional representations is dual to a
membrane ending on this surface \cite{Mal98, Corrado}. The M5-branes
dual to straight and spherical half-BPS Wilson surfaces in higher
dimensional representation were found in \cite{Chen} by solving the
covariant equations of motions for M5-branes\footnote{Similar
M$5$-brane configurations for straight Wilson surface are discussed
in \cite{Lunin07} in Pasti-Sorokin-Tonin (PST) formalism
\cite{Sorokin9701, Sorokin9711} as well. The self-dual string
soliton in $AdS_4\times S^7$ spacetime is discussed in
\cite{Lunin07,Chen2}.}. Analogues to the Wilson loop case, the
worldvolume of the M5-brane dual to the Wilson surface in symmetric
representation has topology $AdS_3\times S^3$ and is completely
embedded in $AdS_7$. While the worldvolume of the M5-brane
corresponding to the Wilson surface in antisymmetric representation
has the same topology but with the $S^3$ part in $S^4$. The
expectation value of the Wilson surfaces in higher dimensional
representation was computed in \cite{Chen} from the action of
M5-brane, without including the subtle boundary terms. The operator
product expansion of Wilson surface operators is computed using
M-theory branes in \cite{Corrado99, ChenLiuWu}.

The Wilson loop or Wilson surface operators are not just probes to
test AdS/CFT correspondence and probe the strings or membranes
dynamics. They are physical gauge invariant observables to
characterize the underlying theory. For example, in pure non-Abelian gauge
theory at finite temperature, the Wilson loop along temporal path,
the so-called Wilson-Polyakov loop, defined by
 \be P(\vec{x})=\frac{1}{N}
{\rm Tr}\left({\cal P}\exp\left(i\int_0^\beta A_0(\vec{x})dt
\right)\right), \ee is the order parameter, characterizing the
phase of the theory. Here $\b=1/T$ is the inverse temperature and
${\cal P}$ denotes the path ordering. Moreover by considering the
correlator of two parallel Wilson-Polyakov loops one can read out
the static potential between two quarks.

 The $AdS_5/CFT_4$
correspondence has a finite temperature extension. At finite
temperature, The spacetime where the four-dimensional field theory lives can be
 either $S^3\times S^1$ or $R^3\times S^1$. Now the time direction becomes
a circle with the period being the inverse of the temperature.
According to the AdS/CFT correspondence, this finite temperature
theory is dual to type IIB string theory on the background which is
the product of a Schwarzschild black hole in $AdS_5$
space\footnote{We denote this by $Sch.$-$AdS_5$.} and a $5$-sphere.
This background comes from the near-horizon limit of non-extremal
black $3$-brane solution of the type IIB supergravity. The Hawking
temperature of the black hole corresponds to the temperature of the
field theory. In \cite{Witten:1998qj, Witten:1998zw}, Witten studied
the thermodynamics of this theory on $S^3\times S^1$. He showed that
there is confinement-deconfinement phase transition in this theory.
This transition is dual to the Hawking-Page transition
\cite{Hawkingpage} in the gravity side. In this case, the AdS/CFT
correspondence tells us  that the Wilson-Polyakov loops in
fundamental representation can be described by fundamental strings
in $Sch.$-$AdS_5$ space \cite{Polyakovloop}. Using this description
the interaction potential between two heavy quarks at finite
temperature was studied. The field theory is on $R^3\times S^1$ in
\cite{Polyakovloop}, so there is no confinement-deconfinement phase
transition. One may expect that for the Wilson-Polyakov loop in
higher representation, D-branes rather than fundamental string are
more appropriate. In \cite{HartnollKumar}, the D5-brane description
of the Wilson-Polyakov loops was proposed. It was also showed in
that paper that there is no D3-brane description. In \cite{Tai}, the
correlation function of two Wilson-Polyakov loops, one in
fundamental representation and the other one in the anti-symmetric
representation, was computed.

Similarly the six-dimensional $(2, 0)$ field theory at finite
temperature is dual to M-theory on $Sch.$-$AdS_7\times S^4$. This
background can come from the the near horizon limit of
non-extermal black M$5$-brane solution. The Hawking temperature in
the gravity side is still corresponding to the temperature in the
field theory side. The gravity dual of this finite temperature
theory was used in \cite{Witten:1998zw} to study the
nonsupersymmtric pure Yang-Mills theory in four dimensions.

In this paper, we would like to study the counterpart of the
Wilson-Polyakov loop in finite temperature six-dimensional $(2,
0)$ field theory in $R^5\times S^1$ using the M-theory branes.
Like the Wilson surface operator, this counterpart should be a
two-dimensional non-local operator. We also expect that it extends
in one spatial direction and one temporal direction. We can still
formally define this operator using eq.~(\ref{ws}), while now
$\Sigma$ should be a surface on which the induced metric has signature
$(1, 1)$. We call this operator Wilson-Polyakov surface.

We propose that when this operator is in lower dimensional
representations it should be described by M2-branes ending on this
surface. We first find the membrane solution corresponding to a
straight Wilson-Polyakov surface. We also compute the potential
between two static parallel self-dual strings with infinite length.
This involves two Wilosn-Polyakov surfaces extending in the same two
directions of the spacetime. The interaction potential between these
two strings can be obtained from the correlation function of these
two Wilson-Polyakov surfaces. There are two classes of membrane
configurations ending on these two Wilson-Polyakov surfaces: one is
of two separated membranes, each of which ending on one
Wilson-Polyakov surface; the other one is a U-shape membrane
connecting these two Wilson-Polyakov surfaces. The interaction
potential is determined by the lowest energy configuration. Denoting
the distance between two strings as $L$ and the temperature as $T$,
 we find that when $LT<<1$, the potential per length goes as $1/L^2$,
similar to the results at zero temperature \cite{Mal98}, and when
$LT>>1$, the interaction potential vanish since it is screened by
the thermal effects.

We also study the M5-brane configurations which should be dual to
the Wilson-Polyakov surfaces in higher dimensional
representations. Similar to the discussions at zero temperature,
two classes of M5-branes are studied. The first class of the
M5-brane is completely embedded in $Sch.$-$AdS_7$, while the
second class of M5-brane has a $S^3$ part embedded in $S^4$. In
the first case, we get a very complicated differential equation.
The existence of the solution of this equation is discussed. While
in the second case, we obtain a class of explicit solutions. Among
these solutions, we indicate a special one which should be dual to
the Wilson-Polyakov surface in the anti-symmetric representation.

Furthermore, inspired by the M5-brane solution dual to the
Wilson-Polyakov surface in antisymmetric representation and the
similar solution at zero temperature in \cite{Chen}, we  consider
M-theory on $M_7\times S^4$ which can be dual to a quite generic
field theory at the boundary of $M_7$. We obtain a universal
result on M5-brane solutions in $M_7\times S^4$. Starting with a
membrane solution whose worldvolume $\Sigma_3$ is completely
embedded in $M_7$, we find that there is always an M5-brane
solution whose topology is $\Sigma_3\times S^3$ with the same
$\Sigma_3$ in $AdS_7$ and $S^3$ in $S^4$. The similar universal
result for D$5$-brane solutions corresponding to Wilson loops in
anti-symmetric representation was discussed in \cite{Hartnoll}.

The investigation we make here may help us to get a better
understanding of the six-dimensional field theory at finite
temperature, the dynamics of the M-theory branes, and even the
dynamics of M-theory itself. For a very good review of the
dynamics of M-theory branes, see \cite{Berman2007}.

The other part of this paper is organized as the following: In
section \ref{m2brane}, we will discuss the M2-brane description of the
Wilson-Polykov surface. Firstly in subsection \ref{straight}, the M2-brane dual
to a straight Wilson-Polykov surface is discussed, then in subsection
\ref{potential}, the interaction between two self-dual strings is studied
using M2-brane description. Our discussions on the M5-brane
description is put in section \ref{m5brane}. In subsection \ref{m5hard}, we investigate
the M5-brane which is completely embedded in $Sch.$-$AdS_7$, in
subsection \ref{schads7s4}, we discuss the M5-brane solution with an $S^3$ part
in $S^4$, and in subsection \ref{universalm5}, we present the universal result we
find. The last section is devoted to conclusion and discussions.

\section{Membrane description\label{m2brane}}

As mentioned before, the six-dimensional $(2, 0)$-field theory at
finite temperature is believed to be dual to the M-theory on
$Sch.$-$AdS_7\times S^4$. We also have a $4$-form flux which fills
in the $4$-sphere. The metric of the background is
 \bea
 ds^2=\frac{R^2}{y^2}(\frac{dy^2}{f(y)}-f(y)dt^2+\sum^5_{i=1}dx^2_i)+\frac{R^2}{4}d\Omega^2_4,
 \label{metric1}
 \eea
 with
 \be
 f(y)=1-\epsilon^6y^6.
 \ee
Here $d\Omega^2_4$ is the metric of unit $4$-sphere. If $\z_i,
i=1, \cdots, 4,$ are the angular coordinates of the $4$-sphere,
then $d\Omega^2_4$ can be written as: \be d\Omega^2_4=
d\z_1^2+\sin^2\z_1 d\z_2^2+\sin^2\z_1 \sin^2\z_2
dz^2_3+\sin^2\z_1^2\sin^2\z_2^2\sin^2\z_3^2 d\z_4^2. \ee The
background $4$-form field strength is \be
H_4=\frac{3R^3}{8}\sin^3\z_1\sin^2\z_2\sin\z_3 d\z_1\w d\z_2 \w
d\z_3\w d\z_4. \ee This background can be obtained from the
near-horizon limit of non-extremal black M$5$-brane solution of
the $11$-dimensional supergravity.  From the AdS/CFT
correspondence, the relation among $R$, $11$-dimensional Plank
length $l_p$ and the parameter $N$ is \be R=(8\pi
N)^{\frac{1}{3}}l_p. \ee In the large N limit, this
six-dimensional field theory at finite temperature should be dual
to the $11$-dimensional supergravity in this background. In the
metric (\ref{metric1}), the boundary field theory is defined at
the conformal infinity where $y=0$, and the coordinates on the
boundary are $t, x_i, i=1, \cdots, 5$. The topology of the
boundary is $R^5\times S^1$ with $S^1$ in the time direction.

Using the standard method, we can get  the Hawking temperature $T_H$
of the black hole: \be T_H=\frac{3}{2\pi} \epsilon. \ee  This
Hawking temperature corresponds to the temperature of the field
theory.

In this paper, we will use AdS/CFT correspondence to study the
Wilson-Polyakov surface operators in the six-dimensional field
theory at finite temperature. In this section, we will study the
M$2$-brane description of the Wilson-Polyakov surface operators.

The bosonic part of the membrane action is\footnote{Our notation
is: the indices from the beginning(middle) of the alphabet refer
to the frame (coordinate) indices, and the underlined indices
refer to the target space ones.} \cite{Bergshoeff:1987cm} \be
S_{M2}=T_2\left(\int d^3\xi\sqrt{-\mbox{det}g_{\mu\nu}}-\int\underline{C}_3\right), \ee
where  $g_{mn}$ is the induced metric on the membrane, $T_2$ is
the tension of M2-brane: \be T_2={1\over (2\pi)^2l_p^3},\ee and
$\underline{C}_3$ is the pullback of the bulk 3-form gauge
potential to the worldvolume of the membrane\footnote{In this
paper, we use the underline indices to denote the target space
indices. We also use the underline to denote the pullback of bulk
gauge potential or field strength to the worldvolume of M2-brane
or M5-brane. We hope that this will not produce confusion. }. The
membrane equations of motions are: \bea
\frac{1}{\sqrt{-g}}\p_m\left(\sqrt{-g}g^{mn}\p_nX^{\underline{N}}\right)G_{\underline{MN}}
+g^{mn}\p_{m}X^{\underline{N}}\p_{n}X^{\underline{P}}\G^{\underline{Q}}_{\underline{NP}}G_{\underline{QM}}
&=&\frac{1}{3!}\epsilon^{mnp}\underline{H}_{\underline{M}mnp}.\nn\\\eea
Here
\underline{H} is the pullback of the background four-form field
strength.

\subsection{Membrane description of Wilson-Polyakov surface\label{straight}}

In this subsection,  we consider a straight Wilson-Polyakov
surface in  six-dimensional $(2, 0)$ theory at finite temperature.
We let it extend in the $t$ and $x_2$ directions. The dual
membrane configuration is always ending on this Wilson-Polyakov
surface. This means that two worldvolume coordinates of membrane
could be identified with $t$ and $x_2$, while the other coordinate
should extend into the bulk.

The simplest membrane configuration is to let it extend only along
$y$ direction. We can easily check that this membrane
configuration is the solutions of the membrane equations of
motion. The only non-trivial equation is the one with index
$\underline{M}=y$, which can be checked by straightforward
calculations. The needed Christoffel symbol of the $Sch.$-$AdS_7$
space is listed in Appendix.

The membrane should be stretched between the boundary ($y=0$) and
the horizon ($y=y_0\equiv 1/\epsilon$) because the non-extremal
$N$ black M$5$-branes should be located at the horizon. Arguments
supporting similar result in the $Sch.$-$AdS_5$ case can be found
in \cite{Maldacena-bh, Polyakovloop}. Some of these arguments can
be applied here. The action of this membrane is \bea
S&=&T_2R^3T_0X_2\int_0^{y_0}{dy \over y^3}\nn\\
&=&\frac{2N}{\pi}T_0X_2\int_0^{y_0}{dy \over y^3}.\eea where $T_0,
X_2$ are the lengths of the $t, x_2$ direction, and in the second
line of the above equation, $T_2R^3=2N/\pi$ is used.
 By introduce a cutoff $y=\delta$ near
$y=0$, we get, \be
S=-\frac{N}{\pi}T_0X_2(\frac{1}{y_0^2}-\frac{1}{\delta^2}). \ee

\subsection{The potential between two static strings\label{potential}}

It would be interesting to study the potential between two static
parallel infinitely-long strings in this six-dimensional
finite-temperature field theory. We let these strings extend along
the $x_2$ direction and put them at $x_1=L/2$ and $x_1=-L/2$. To
compute this potential, we need to consider two Wilson-Polyakov
surfaces. These surfaces should extend along the $t$ and $x_2$
direction and be put at $x_1=L/2$ and $x_1=-L/2$, respectively.
\footnote{We choose the same scalar coupling for these two
Wilson-Polyakov surfaces, in another word, we choose the same
point at $S^4$ for them.}

There are two classes of membrane configurations ending on these two
Wilson-Polyakov surfaces. The first is two separated parallel
membranes, with each membrane ending on one of the two
Wilson-Polyakov surfaces and extending along $y$ direction to the
horizon. These membranes and their actions have been studied in the
last subsection. The second is a U-shape membrane connecting these
two Wilson-Polyakov surfaces. Now we will study this membrane
solution. In the large $N$ limit, the potential between these two
static strings will be determined by the lowest energy membrane
configuration among the possible classical solutions.

The connected membrane solution will only extend in $x_1, x_2, t,
y$ directions of the background geometry. We choose the
coordinates of the worldvolume of the corresponding membrane to be
$x_1, x_2, t$, and $y$ is a function of $x_1$ only. Then boundary
condition is: $y(-L/2)=y(L/2)=0$.

The induced metric on this membrane is \bea
ds^2_{\mbox{ind}}=\frac{R^2}{y^2}\left(-fdt^2+dx^2_2+(1+\frac{y'^2}{f})dx^2_1\right),
\eea where $y'$ is $dy/dx_1$. Then the action of this membrane is,
\bea S=T_2\int dtdx_1dx_2 \sqrt{-g}=T_2R^3T_0X_2\int
\frac{\sqrt{f+y'^2}}{y^3}dx_1,\eea where $T_0, X_2$ are the
lengths of the $t, x_2$ direction.

\begin{figure}[h!]
    \epsfxsize=100mm%
    \hfill\epsfbox{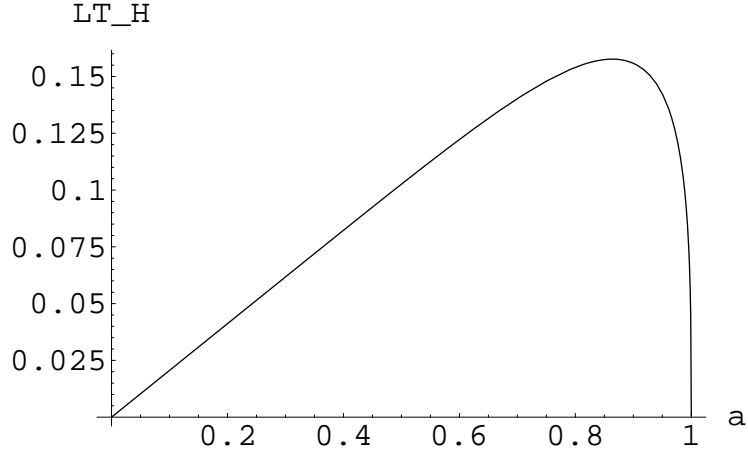}\hfill~\\
    \caption{The functional relation of $LT_H$ and $a$. One can see that
    for $L<L_{max}$, there are two connected membrane solutions, while for $L>L_{max}$,
    there are no connected   membrane solutions.}
    \label{figLT}
   \end{figure}

We can consider the above action as the one of an imaginary particle
with $x_1$ plays the role of time. Since the Lagrangian does not
depend on $x_1$ explicitly, the Hamiltonian in the $x_1$ direction
is a constant of motion. So \bea p_y y'-{\cal
L}=-\frac{T_2R^3T_0X_2f}{y^3\sqrt{f+y'^2}}=const. \eea According to
the symmetries of this system, at $x_1=0$, $y$ should reach its
maximum value $y_m$. So at this point, $y'=0$. Note that we only
consider the membrane solution out of the horizon, so $y_m\leq y_0$.
Using the above equation, we have \bea
\frac{f(y)}{y^3\sqrt{f(y)+y'^2}}=\frac{f(y_m)}{y_m^3\sqrt{f(y_m)}}=\frac{\sqrt{f(y_m)}}{y_m^3},
\eea so \be y'^2=\frac{(1-\r06 y^6)(y_m^6-y^6)}{y^6(1-\r06y_m^6)}.
\ee We can solve this equation to get: \be x_1=\left\{
\begin{array}{cc}
\int^{y_m}_y\sqrt{\frac{1-\r06 y_m^6}{(1-\r06
z^6)(y_m^6-z^6)}}z^3dz, & \mbox{when $x_1>0$;}\\
-\int^{y_m}_y\sqrt{\frac{1-\r06 y_m^6}{(1-\r06
z^6)(y_m^6-z^6)}}z^3dz, & \mbox{when $x_1<0$.}
\end{array}
\right. \end{equation} Because of the boundary conditions, we have
the following relations between $y_m$ and $L$: \bea
L&=&2\int^{y_m}_0\sqrt{\frac{1-\r06 y_m^6}{(1-\r06
z^6)(y_m^6-z^6)}}z^3dz. \eea By introducing $a\equiv \epsilon
y_m$, $L$ can be written as: \bea L&=&{2a\over \epsilon}
\sqrt{1-a^6}\int_0^1{z^3dz\over \sqrt{(1-z^6)(1-a^6z^6)}}\eea
The above result can be expressed by the hypergeometric function
as the following: \be L={2\sqrt{\pi}\Gamma(2/3)a\over
\epsilon\Gamma(1/6)} \sqrt{1-a^6}\, {}_2F_1(\frac{1}{2},
\frac{2}{3}, \frac{7}{6}, a^6). \ee
From this, we can also get the dimensionless combination $LT_H$ as a
function of $a$:
\begin{equation}
LT_H={3\Gamma(2/3)a\over \sqrt{\pi}\Gamma(1/6)} \sqrt{1-a^6}\,
{}_2F_1(\frac{1}{2}, \frac{2}{3}, \frac{7}{6}, a^6).
\label{LT}\end{equation} This function is plotted in
Fig.~\ref{figLT}.

One can see that $L$ has a maximal value $L_{\mbox{max.}}$. For
each $L$ less than $L_{\mbox{max.}}$, there are two corresponding
$a$'s: $a_1(L)$ and $a_2(L)$ with $a_1(L)<a_2(L)$. So there are
two connected membrane configurations. For $L>L_{\mbox{max.}}$,
there are no connected membrane solutions.

For the connected membrane solution, the action is: \be
S^{\mbox{\small
con.}}(a)=2T_2R^3T_0X_2\int_0^{y_m}\frac{1}{y^3}\sqrt{\frac{y_m^6(1-\r06
y^6)}{y_m^6-y^6}}dy. \ee As in \cite{Mal97, Polyakovloop}, We
should subtract the action of two straight membranes stretched
between the boundary ($y=0$) and the horizon ($y=y_0\equiv
1/\epsilon$).\footnote{This action represents the masses of these
two strings.} These membranes extend in the $y, t, x_2$ directions
and are just the disconnected membrane configuration. After the
subtraction, the action is \bea S^{\mbox{\small con.}}_{\mbox{\it
ren.}}(a)&=&2T_2R^3T_0X_2\left(\int_0^{y_m}\frac{1}{y^3}\sqrt{\frac{y_m^6(1-\r06
y^6)}{y_m^6-y^6}}dy-\int_0^{y_0}{dy \over y^3}\right)\nn\\
&=&\frac{4NT_0X_2}{\pi}
\epsilon^2\left[\frac{1}{a^2}\int_0^1\left(\frac{1}{z^3}\sqrt{\frac{1-a^6z^6}{1-z^6}}-\frac{1}{z^3}\right)dz
+\frac{1}{2}\left(1-\frac{1}{a^2}\right)\right]. \nn\\
&=& \frac{2NT_0X_2}{\pi}
\epsilon^2\left(1+\frac{\sqrt{\pi}\Gamma(-1/3)}{3a^2\Gamma(1/6)}\, {}_2F_1(-\frac{1}{2},
-\frac{1}{3}, \frac{1}{6}, a^2)\right)\eea

\begin{figure}[h!]
    \epsfxsize=100mm%
    \hfill\epsfbox{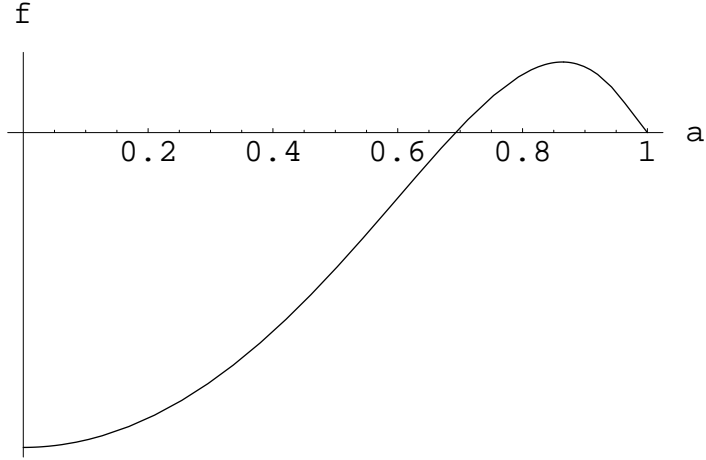}\hfill~\\
    \caption{$f\equiv S^{\mbox{\small con.}}_{\mbox{\it ren.}}L^2/(T_0X_2)$ as a function of $a$. One can see that for
    $a<a_c$, the renormalized action is negative; while for $a>a_c$, the renormalized action is positive.}
    \label{figenergy}
       \end{figure}

 The dimensionless
combination $S^{\mbox{\small con.}}_{\mbox{\it ren.}}L^2/(T_0X_2)$
is the following function of $a$:
\begin{equation}
S^{\mbox{\small con.}}_{\mbox{\it ren.}}L^2/(T_0X_2)=\frac{8\pi
N}{9}(LT_H)^2
\left(1+\frac{\sqrt{\pi}\Gamma(-1/3)}{3a^2\Gamma(1/6)}\, {}_2F_1(-\frac{1}{2},
-\frac{1}{3}, \frac{1}{6}, a^2)\right),\label{energy}
\end{equation}
where $(LT_H)^2$ is given in eq.~(\ref{LT}).
This function is plotted in Fig.~\ref{figenergy}.

\begin{figure}[h!]
    \epsfxsize=100mm%
    \hfill\epsfbox{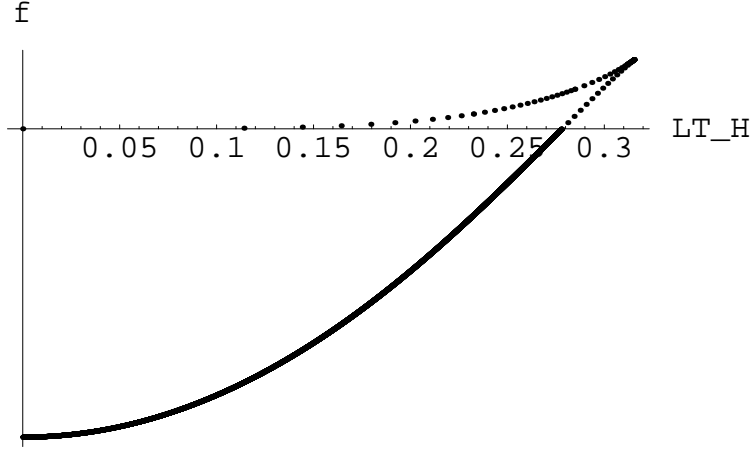}\hfill~\\
    \caption{The functional relation between $f\equiv S^{\mbox{\small con.}}_{\mbox{\it ren.}}(a)L^2/(T_0X_2)$ and $LT_H$.
    The upper dashed curve is the one corresponding to the membrane configuration with the smaller $a$, i. e., $a_2$;
    The lower solid-dashed curve is the one corresponding to the membrane configuration with the larger $a$, i. e., $a_1$.
    For $L<L_c$, the solid curve gives us $VL^2/X_2$ as a function of $LT_H$, where $V$ is the interaction potential per length
    between these two strings, while for $L>L_c$, the interaction potential is zero. We note that this result is at the leading order of large
    $N$ expansion. }
    \label{figLTenergy}
   \end{figure}

After eliminating $a$, we can obtain the functional relation
between these two dimensionless combinations: $S^{\mbox{\small
con.}}_{\mbox{\it ren.}}L^2/(T_0X_2)$ and $LT_H$. This functional
relation is plotted in Fig.~\ref{figLTenergy}.

As to the disconnected membrane configuration, this solution
always exists for any $L$. Due to our substraction prescription,
the renormalized action vanishes: \be S^{\mbox{\small
con.}}_{\mbox{\it ren.}}(a)=0. \ee

One can see from Fig.~\ref{figenergy} that there is a value $a_c
\approx 0.692$, such that $S^{\mbox{\small con.}}_{\mbox{\it
ren.}}(a_c)=0$. One can also see that \be S^{\mbox{\small
con.}}_{\mbox{\it ren.}}(a)\left\{\begin{array}{cc} <0, &
\mbox{when $a<a_c$;}\\>0, & \mbox{when $a>a_c$}.
\end{array}\right.
 \ee We can also find from Fig.~\ref{figLT} that for any
$L<L_{\mbox{max.}}$, $a_2(L)>a_c$.

So for any given $L<L_{\mbox{max.}}$, among the renormalized
action of three possible membrane configurations, $S^{\mbox{\small
con.}}_{\mbox{\it ren.}}(a_1(L))$, $S^{\mbox{\small
con.}}_{\mbox{\it ren.}}(a_2(L))$, and $S^{\mbox{\small
dis.}}_{\mbox{\it ren.}}(L)$, the smallest one
is\footnote{Fig.~\ref{figLTenergy} tells us that for any
$L<L_{max}$, we always have $S^{\mbox{\small con.}}_{\mbox{\it
ren.}}(a_1(L))<S^{\mbox{\small con.}}_{\mbox{\it ren.}}(a_2(L))$.}
\be \left\{\begin{array}{cc}
S^{\mbox{\small con.}}_{\mbox{\it ren.}}(a_1(L)), & \mbox{when $a\le a_c$;}\\
S^{\mbox{\small dis.}}_{\mbox{\it ren.}}(L)=0,& \mbox{when $a\ge
a_c$.}\end{array}\right.\ee While for any $L>L_{\mbox{max.}}$, the
only possible membrane configuration is the disconnected one whose
renormalized action vanishes.

So at the leading order of large $N$ expansion, the interaction
potential per length between two infinite strings is: \be
\frac{V}{X_2}=\left\{
\begin{array}{cc} S^{\mbox{\small con.}}_{\mbox{\it ren.}}(a_1(L))/(X_2T_0), & \mbox{for $L\le
L_c$;}\\ 0 & \mbox{for $L\ge L_c$.}\end{array}\right.\ee  Here
$L_c$ ($\approx 0.278/T_H$) is the value of $L$ such that
$a_1(L_c)=a_c$. One can see from Fig.~\ref{figLT} and
Fig.~\ref{figenergy} that $L_c<L_{\mbox{max.}}$. For $L<L_c$, the
functional relation between the dimensionless combination
$VL^2/X_2$ and $LT_H$ is given by the solid curve in
Fig.~\ref{figLTenergy}. Physically, we can take $L_c$ as the
screening length. When $L\leq L_c$, the two strings can interact
with each other. And when $L\geq L_c$, the two strings are
screened by the thermal fluctuation and de-associate.

Now we further study the potential in the case of $LT_H<<1$. In
this case, we have $a<<1$. Under this condition, we can expand
Eq.~(\ref{LT}) in powers of $a$, \be
LT_H=\frac{3c}{\pi}a\left(1-\frac{3}{14}a^6-\frac{75}{728}a^{12}+\cdots
\right), \ee where \be c\equiv\frac{\sqrt{\pi}\Gamma({2\over
3})}{\Gamma({1\over 6})}.\ee Then we can get \be a=t
(1+\frac{3}{14}t^6+\frac{309}{728}t^{12}+\cdots), \ee where
$t\equiv \pi L T_H/(3c)$.

From this result and the expansion of eq.~(\ref{energy}) in powers of $a$, we get
\be
\frac{V}{X_2}=\frac{N}{L^2}\left(-\frac{8c^3}{\pi}+\frac{8\pi}{9}(LT_H)^2
-\frac{32\pi^5}{5103c^3}(LT_H)^6+\cdots\right).
 \ee

Take $T_H\to 0$ in above equation, we arrive at the zero temperature result in \cite{Mal98}: \be
V/X_2=-\frac{8\sqrt{\pi}\Gamma(\frac{2}{3})^3N}{\Gamma(\frac{1}{6})^3L^2},
\ee which is always lower than the finite temperature results.

In summary, the asymptotic behavior of our results at finite
temperature is: when $L<<1/T_H$, $V/X_2$ goes like $1/L^2$ similar
to what happens at zero temperature; while when $L>>1/T_H$, the
potential is zero since the interaction is screened by the finite
temperature effects.

From the above discussion, one can also see that $VL^2/X_2$
depends on $T_H$ only through the combination $T_HL$, this is due
to the underlying conformal symmetry although this symmetry is
broken at finite temperature.

\section{M5-brane description\label{m5brane}}

In this section, we turn to study the M$5$-brane description of
the straight Wilson-Polyakov surface operator. We expect that this
description should be a better one when the Wilson-Polyakov
surface is in higher dimensional representations, like what
happens in the zero-temperature case \cite{Chen}.

Let us first give a brief review of the  M$5$-brane covariant
equations of motion in an eleven-dimensional curved
spacetime\cite{Sezgin97}. We are only interested in the bosonic
components of the equations, which include the scalar equation and
the tensor equation. The scalar equation takes the form
 \be\label{scalareq}
 G^{mn}\nabla_m \cE_n^{\underline
 c}=\frac{Q}{\sqrt{-g}}\epsilon^{m_1\cdots
 m_6}\big(\frac{1}{6!}H^{\underline a}_{~m_1\cdots
 m_6}+\frac{1}{(3!)^2}H^{\underline
 a}_{~m_1m_2m_3}H_{m_4m_5m_6}\big)P_{\underline
 a}^{~\underline c}
 \ee
 and the tensor equation is of the form
 \be\label{tensoreq}
 G^{mn}\nabla_mH_{npq}=Q^{-1}(4Y-2(mY+Ym)+mYm)_{pq}.
 \ee

The various quantities in the above equations of motion are
introduced as follows.  There exist a self-dual 3-form field
strength $h_{mnp}$ on the M5-brane worldvolume, from which, one
can define
 \bea
 k_m^{~n}&=&h_{mpq}h^{npq}, \\
 Q&=&1-\frac{2}{3}\Tr k^2, \\
 m_p^{~q}&=&\delta_p^{~q}-2k_p^{~q}, \\
 H_{mnp}&=&4Q^{-1}(1+2k)_m^{~q}h_{qnp}
 \eea
 Note that $h_{mnp}$ is self-dual with respect to worldvolume
 metric but not $H_{mnp}$, which instead satisfies a nonlinearly self-dual
 condition and also the Bianchi
 identity
 \be
 dH_3=-{\underline H}_4
 \ee
 where ${\underline H}_4$ is the pull-back of the target space 4-form flux. The
 induced metric is simply
 \be
 g_{mn}=\cE_m^{\underline a}\cE_n^{\underline b}\eta_{\underline
 ab}
 \ee
 where
 \be
 \cE_m^{\underline a}=\p_mz^{\underline m}E_{\underline m}^{\underline
 a}.
 \ee
Here $z^{\underline m}$ is the target spacetime coordinate, which
is a function of worldvolume coordinate $\xi$ through embedding,
and $E_{\underline m}^{\underline
 a}$ is the component of target space vielbein. However, it is not $g_{mn}$ but instead
 another tensor
 \be\label{Gmn}
 G^{mn}=(1+\frac{2}{3}k^2)g^{mn}-4k^{mn},
 \ee
 which appear in (\ref{scalareq}). And the covariant derivative in
 (\ref{scalareq}) means
  \be
 \nabla_m\cE_n^{\underline c}=\p_m\cE_n^{\underline c}-\G^p_{mn}\cE_p^{\underline
 c}+\cE_m^{\underline a}\cE_n^{\underline b}\o^{\underline c}_{{\underline a}{\underline
 b}}
 \ee
 where $\G^p_{mn}$ is the Christoffel symbol with respect to the induced worldvolume
 metric and $\o^{\underline c}_{{\underline a}{\underline
 b}}$ is the spin connection of the background spacetime.
 Also one has
 \be
 P_{\underline a}^{~\underline c}=\delta^{\underline
 c}_{\underline a}-\cE_{\underline a}^m\cE_m^{~{\underline c}}.
 \ee

Moreover, there is a 4-form field strength $H_{{\underline
a}_1\cdots {\underline a}_4}$ and its Hodge dual
 7-form field strength $H_{{\underline a}_1\cdots {\underline
 a}_7}$:
 \bea
 H_4&=&dC_3 \nn\\
 H_7&=&dC_6+\frac{1}{2}C_3\w H_4
 \eea
 The frame indices on $H_4$ and $H_7$ in the scalar and the tensor equations have
 been converted to worldvolume indices with factors of
 $\cE_m^{\underline c}$.
  From them, we can define
 \be
 Y_{mn}=[4\star {\underline H}-2(m\star {\underline H}+\star {\underline H}m)+m\star {\underline H}m]_{mn},
 \ee
where \be \star {\underline
H}^{mn}=\frac{1}{4!\sqrt{-g}}\epsilon^{mnpqrs}{\underline
H}_{pqrs}. \ee These two quantities appear in the tensor equation
of motion.

These equations of motion can be obtained from the non-chiral
action \cite{Sundell97, Sezgin99} or the PST (Pasti-Sorokin-Tonin)
action \cite{Sorokin9701, Sorokin9711}. 
In the non-chiral action, a nonlinear self-dual condition for $H_3$
should be put by hand instead of coming from the variation of the
action. This is similar to what happens in the case of
ten-dimensional type IIB supergravity where the self-dual condition
for $5$-form field strength is put by hand. In the PST action, an
auxiliary field is introduced to deal with the self-duality of
$H_3$. We postpone a brief introduction of the PST action to the
subsection \ref{universalm5}, since only there this action is
needed.

\subsection{M5-brane configuration in $Sch.$-$AdS_7$\label{m5hard}}

First we consider the M5-brane solution which is completely
embedded in the $Sch.$-$AdS_7$ part of the background metric. In
this case, we expect that due to the membrane interaction in the
presence of background 4-form flux, the membrane will polarize to
a M5-brane by blowing up an $S^3$ in the transverse direction.
This is really the case for the Wilson surface operators discussed
in \cite{Chen}. Now we choose the coordinates of $Sch.$-$AdS_7$
such that the metric takes the following form: \be
ds^2=\frac{R^2}{y^2}(-fdt^2+\frac{dy^2}{f}+dx^2+dr^2+r^2d\Omega^2_3).
\ee

 In the case of the
straight Wilson-Polyakov surface, let the worldvolume coordinates
of M5-brane be $\xi_i,~~ i=0,\cdots, 5$, and the embedding be
 \bea
 \xi_0=t,~~~ \xi_1=x,~~~ \xi_2=y,~~~ r=g(y), \\
 \xi_3=\a,~~~ \xi_4=\b, ~~~ \xi_5=\g,
 \eea
 where $\a,\b,\g$ are the angular coordinates of $S^3$. This embedding
 is reasonable from the experience in the study of the Wilson surface operators. The
 induced metric is then
 \bea
 ds^2_{\mbox{ind}}&=&\frac{R^2}{y^2}(-fd\xi_0^2+d\xi_1^2+(f^{-1}+g^{\pr
 2})d\xi_2^2+g^2d\O_3^2)\nn\\
 &=&\frac{R^2}{y^2}(-fdt^2+dx^2+(f^{-1}+g^{\pr
 2})dr^2)+\frac{R^2g^2}{y^2}(d\a^2+\sin^2\a d\b^2+\sin^2\a
 \sin^2\b d\g^2)\nn\\
 & &\label{indmetric}
 \eea
 where the prime denotes the derivative with respect to $y$. Without causing confusion, we
 simply let $t,x,y,\a,\b,\g$ be the
 coordinates of the M5-brane worldvolume.

There is a self-dual 3-form field strength in the M5-brane
worldvolume. Let us assume it to be
 \be
 h_3=\frac{a}{2}(1+\star_{\mbox{ind}})\sqrt{\det G}d\a \wedge d\b
 \wedge d\g
 \ee
 where $a$ could be a function of $y$ and $\det G$ is the determinant of the metric of $S^3$. In our
 case, we have
 \be
 h_3=\frac{a}{2}(\frac{R}{y})^3(g^3\sin^2\a\sin\b d\a\w d\b\w
 d\g+\sqrt{1+fg^{\pr 2}}dt\w dx\w dy).
 \ee

Then we can calculate the relevant quantities $k^{mn}, G^{mn}$
etc.. It turns out that the physical 3-form field strength is
 \be
 H_3=2a(\frac{R}{y})^3(\frac{\sqrt{1+fg^{\pr
 2}}}{1+a^2}dt\w dx \w dy +\frac{g^3}{1-a^2}\sin^2\a\sin\b d\a \w
 d\b\w d\g)
 \ee

Since there is no pull-back of bulk 4-form field strength on the
 M5-brane worldvolume, we have $dH_3=0$, which gives the
 constraint
 \be\label{cons}
 \frac{a}{1-a^2}\frac{g^3}{y^3}=\mbox{constant}
 \ee

The equation of motion on the tensor $H_{npq}$, in this case, is
 \be
 G^{mn}\nabla_m H_{npq}=0.
 \ee
Here $\nabla_m$ is the covariant derivative with respect to the
induced metric. We list the detailed Levi-Civita connection in
Appendix. It is somehow surprising that the tensor equation give the
same constraint (\ref{cons}).  It is remarkable that (\ref{cons}) is
independent of the form of $f$.

For the scalar equation of motion, it is more involved. In our
case, we have
 \bea
 \cE^{\underline 0}_t=\frac{R}{y}\sqrt{f}, ~~~ \cE^{\underline 1}_{x}=\frac{R}{y}, ~~~\cE^{\underline
 2}_{y}=\frac{R}{y\sqrt{f}}, ~~~ \cE^{\underline 3}_{y}=\frac{R}{y}g^\prime,\nn\\
 \cE^{\underline 4}_{\a}=\frac{Rg}{y},~~~ \cE^{\underline
 5}_{\b}=\frac{Rg\sin\a}{y}, ~~~ \cE^{\underline
 6}_{\g}=\frac{Rg\sin\a\sin\b}{y},
 \eea
where we have set the veilbein of $AdS_7$ part of the target
spacetime as
 \bea
 \hat{\th}^0=\frac{R}{y}\sqrt{f}dt,~~~\hat{\th}^1=\frac{R}{y}dx,~~~\hat{\th}^2=\frac{R}{y\sqrt{f}}dy,~~~\hat{\th}^3=\frac{R}{y}dr,\nn\\
 \hat{\th}^4=\frac{Rr}{y}d\a,~~~\hat{\th}^5=\frac{Rr\sin\a}{y}d\b,~~~\hat{\th}^6=\frac{Rr\sin\a\sin\b}{y}d\g.
 \eea
The corresponding spin connection could be found in Appendix.
 The straightforward
 calculation shows that
 \be
 G^{mn}\nabla_m\cE_n^{\underline c}=0,\hspace{5ex} \mbox{except ${\underline
 c}={\underline 2}$ or ${\underline 3}$}. \ee
 The nontrivial components come from ${\underline c}={\underline 2}$ or ${\underline
 3}$. The right hand side of the scalar equation of motion consists of
 the matrix $P^{\underline c}_{\underline a}=\delta^{\underline
 c}_{\underline a}-\cE_{\underline a}^m\cE_m^{~{\underline c}}$,
 which has nonvanishing components
  \be
 P^{~\underline c}_{\underline a}=\left(\begin{array}{cc}
 \frac{fg^{\prime 2}}{1+fg^{\prime 2}}&-\frac{\sqrt{f}g^{\prime}}{1+fg^{\prime 2}}\\
 -\frac{\sqrt{f}g^{\prime}}{1+fg^{\prime 2}}&\frac{1}{1+fg^{\prime 2}}
 \end{array}\right).
 \ee
 where ${\underline a},{\underline c}$ take values ${\underline 2},{\underline
 3}$.

 For the background flux, we have a dual 7-form field strength in
 $AdS_7$ part,
  \be
  H_{{\underline 0}{\underline 1}\cdots {\underline
  6}}=\frac{6}{R}
  \ee
Note that our convention is a little different from the literature
by a factor $2$ since we have rescaled the radius of $AdS_7$. On
the right hand side of the scalar equation, only 7-form field
strength contributes since the M5-brane worldvolume is embedded
simply into $AdS_7$ and there is no induced 4-form field strength
on it.

 It turns out that the nontrivial components ${\underline c}={\underline 2}$ and ${\underline
 3}$ of the scalar equation of motion give the same constraint:
  \bea
  \frac{6(1-a^4)}{\sqrt{1+fg^{\prime
  2}}}&=&(1+a^2)^2\left\{\frac{3fg^\prime}{1+fg^{\prime
  2}}-\frac{1}{(1+fg^{\prime
  2})^2}(\frac{yf^\prime g^\prime}{2}(2+fg^{\prime
  2})+fyg^{\prime\prime})\right\}\nn\\
  & &+3(1-a^2)^2\frac{1}{1+fg^{\prime
  2}}(fg^\prime+\frac{y}{g}).\label{scalar}
 \eea
When one takes $f=1$ and $g=\k^{-1}y$, the above equation is just
the one for the Wilson surface operator in the symmetric
representation, which was discussed in \cite{Chen}. Generically even
when one takes $f=1$, the equation (\ref{scalar}) is quite hard to
solve analytically. When one consider the $Sch.$-$AdS_7$ with a
nonconstant $f$, even the existence of the solution is not an easy
problem. In \cite{HartnollKumar}, it was showed that there are no
D3-brane solutions with finite total action dual to Wilson-Polyakov
loops in four dimensional ${\cal N}=4$ super Yang-Mills theory at
finite temperature. In the case at hand, we can not directly use
their argument since the total action of M5-brane is still not
well-defined due to the subtlety of the boundary terms and the
conformal anomalies\cite{Witten99}.  Here we would like to just
discuss the existence of the solution of the above differential
equation.  Let us impose the following initial condition: \be
g(0)=c_1, \hspace{5ex}g'(0)=c_2. \ee  For the case of $c_1=0$, we
have mentioned that this initial value problem has a solution
$g=\k^{-1}y$ when $\epsilon=0$. This will guarantee that for small
enough (positive) $\epsilon$, the above initial value problem will
have a solution in a finite interval $[0, y_0(\epsilon)]$. In
another word, when the temperature is low enough, the M5-brane
solution dual to Wilson surface in symmetric representation in
zero-temperature theory will only be deformed, not be destroyed.
However, for the case of $c_1\ne 0$, we find that this initial value
problem has no solutions.\footnote{We would like to thank Antonio
Ambrosetti and Jiayu Li for discussions and helps on the study of
this ordinary differential equation.}

\subsection{M5-brane configuration in $Sch.$-$AdS_7 \times S^4$\label{schads7s4}}

Now let us consider another possibility. We consider the
M$5$-brane solution with topology $\Sigma_3\times S^3$. Now
$\Sigma_3$ will be in $Sch.$-$AdS_7$ and $S^3$ in $S^4$. Let the
worldvolume coordinates of M5-branes be $\xi_i$, $i=0,\cdots 5$
and the embedding be
 \bea
 \xi_0=t,~~ \xi_1=x, ~~\xi_2=y,~~r=g(y) \nn\\
 \xi_3=\z_2, ~~\xi_4=\z_3,~~\xi_5=\z_4,~~ \z_1=\z^0
 \label{m5brane2}\eea
 where $\z_i$ are the angular coordinates of $S^4$. Here we let
 $\z_1$ be fixed at a constant $\z^0$. The induced metric is
 \bea\label{indmetric3}
 ds^2_{\mbox{ind}}=\frac{R^2}{y^2}(-fdt^2+dx^2+(f^{-1}+g^{\pr 2})dy^2)+\frac{R^2\sin^2\z^0}{4}(d\z_2^2
 +\sin^2\z_2d\z_3^2+\sin^2\z_2\sin^2\z_3d\z^2_4).\nn\\
 \eea

In this case, we take the self-dual 3-form field strength on the
M5-brane worldvolume to be \bea
 h_3=\frac{1}{2}aR^3(\frac{\sqrt{1+fg^{\pr 2}}}{y^3}dt\w dx \w dy+
 \frac{\sin^3\z^0}{8}\sin^2\z_2 \sin\z_3 d\z_2\w d\z_3 \w d\z_4)
 \eea
where $a$ could be a function of $y$.

Similar to the above cases, we can get $k^{mn}$,
$k^2=\frac{3}{2}a^4$ and  $Q=1-a^4$. And the physical 3-form is
 \bea
 H_3=2aR^3(\frac{\sqrt{1+fg^{\pr 2}}}{(1+a^2)y^3}dt\w dx \w dy+
 \frac{\sin^3\z^0}{8(1-a^2)}\sin^2\z_2 \sin\z_3 d\z_2\w d\z_3 \w
 d\z_4).\label{eqH3}
 \eea
The condition that $dH_3=0$ requires that $a$ is a constant.

It is straightforward to check if it is possible and under what
condition if possible that the above ansatz satisfy the equations of
motion. Since $a$ is a constant, the tensor equation is satisfied.
And from the scalar equation, for the trivial embedding in $AdS_7$
$r=\mbox{constant}$ and the nontrivial embedding in $S^4$, the
discussion is parallel to the one in \cite{Chen}, we get \be
\label{az}a=\frac{\pm 1+\sin\zeta^0}{\cos\zeta^0}. \ee As for the
nontrivial embedding in $Sch.$-$AdS^7$ part, it is somehow
interesting. Firstly note that the R.H.S of scalar equation is
always vanishing in this case due to the pull-back of the 4-form or
dual 7-form field strength is zero. At the end, we have the
following equation:
 \be
 \frac{3fg^\prime}{1+fg^{\prime
  2}}-\frac{1}{(1+fg^{\prime
  2})^2}(\frac{yf^\prime g^\prime}{2}(2+fg^{\prime
  2})+fyg^{\prime\prime})=0,
 \ee
which can be cast into the form
 \be \label{eg}
 g^{\pr\pr}+\frac{f^\pr g^\pr}{2f}(2+fg^{\pr
 2})-\frac{3}{y}g^\pr(1+fg^{\pr 2})=0.
 \ee
Obviously when $g$ is a constant, which means that the embedding
in $AdS_7$ is trivial, the above equation is satisfied, no matter
what $f$ is. This means that in this case we always have a
M5-solution once (\ref{az}) holds, just as we expected.  We
propose here that the solution with $g=0$ should be dual to a
straight Wilson-Polyakov surface operator in higher dimensional
antisymmetric representation.

Certainly it would be interesting to solve Eq. (\ref{eg}). It looks
simpler than the one for the symmetric case, but still hard to
solve. For example, let $f=1$,  which reduce to the background
without the Schwarzschild blackhole. The equation is reduced to
 \be
 g^{\pr \pr}-\frac{3}{y}g^\pr(1+g^{\pr 2})=0.
 \ee
It could be solved exactly: \bea
g&=&c_1-\frac{1}{2}c_0^{-1/3}\left((3^{-1/4}-3^{1/4})F(\beta,
\frac{1+\sqrt{3}}{2\sqrt{2}})\right.\nn\\
&+&\left.2\sqrt[4]{3}E(\beta,
\frac{1+\sqrt{3}}{2\sqrt{2}})-\frac{2\sqrt{1-c_0^2y^6}}{\sqrt{3}+1-c_0^{1/2}y^3}
)\right), \eea where $c_0$ and $c_1$ are two integral constants
with $c_0$ being non-negative,  $\beta$ is defined as \be
\beta=\arccos\frac{\sqrt{3}-1+c_0^{2/3}y^2}{\sqrt{3}+1-c_0^{2/3}y^2},\ee
and  $F$ and $E$ are elliptic integrals of the first and second
kind, respectively. In this solution $y$ can only take the value
between $0$ and $c_0^{-1/3}$.
 Obviously $g$ being a constant is a trivial
embedding. And the special one with $g=0$ corresponds to the
Wilson surface operator in anti-symmetric representation. However
it is remarkable that for the pure $AdS_7\times S^4$ case, there
actually exist a two-parameter class of M5-brane configuration,
characterized by the integral constant $c_0,c_1$. The one with $g$
being constant is the one with half supersymmetries. However, with
$f$ not being a constant, the equation (\ref{eg}) is hard to
solve.

The key point in the above discussion is that the embeddings in
$AdS_7$ and $S^4$ are independent.

\subsection{A universal result\label{universalm5}}

As a generalization of the M$5$-brane solutions corresponding to
Wilson(-Polyakov) surfaces in antisymmetric representation found
in \cite{Chen} and the previous subsection, we will prove a
universal result on a class of  M$5$-brane solutions in this
subsection. We consider M-theory on $M_7\times S^4$ with four form
fluxes filling in $S^4$. We assume that this background is the
solution of the eleven dimensional supegravity and a good
background of M-theory. If M-theory on this background is dual to
a field theory on the boundary of $M_7$, we expect this universal
result is useful to study the Wilson(-Polyakov) surface operators
in the field theory on the boundary. We need not to require that
this background has any supersymmetries. $AdS_7$ and
$Sch.$-$AdS_7$ are two special examples of $M_7$.

The background metric on $M_7\times S^4$ is
\begin{equation}
 ds^2_{M_7\times S^4}=ds^2_{M_7}+\frac{R^2}{4}\left(d\z_1^2+\sin^2\z_1 d\z_2^2+\sin^2\z_1
\sin^2\z_2 dz^2_3+\sin^2\z_1^2\sin^2\z_2^2\sin^2\z_3^2 d\z_4^2
\right).
\end{equation}

We assume that there is a membrane solution in this background and
the worldvolume of this membrane, $\Sigma_3$, is completely
embedded in $M_7$ part of the background geometry. Locally we can
always choose the coordinates of the worldvolume such that the
induced metric takes the following diagonal form:
\begin{equation}
 ds^2_{\Sigma_3}=g_{\xi_0\xi_0}d\xi_0 d\xi_0+g_{\xi_1\xi_1}d\xi_1 d\xi_1+g_{\xi_2\xi_2}d\xi_2 d\xi_2.
\end{equation}
This worldvolume is a three-dimensional submanifold of $M_7$ with
minimal volume.

Now, we plan to show that from this membrane solution, we can
obtained a M5-brane solution whose worldvolume has topology
$\Sigma_3\times \tilde S^3$ with the same $\Sigma_3$ in $M_7$ and
$\tilde S^3$ in $S^4$.

Since here $M_7$ is quite generic, it is not easy to search for
the M$5$-brane solution using the covariant M$5$-brane equations
of motion. So in our discussions here we will use the PST
(Pasti-Sorokin-Tonin) action \cite{Sorokin9701, Sorokin9711} of
the M5-brane as in \cite{Lunin07}. The bosonic part of the PST action is the
following:
\begin{equation}
 S_{PST}=T_5\int d^6x\left(\sqrt{-\mbox{det}(g_{mn}+i\tilde H_{mn})}-\frac{\sqrt{-g}}{4}\tilde H^{mn}H_{mn}\right)
-T_5\int Z_6,
\end{equation}
where
\begin{equation}
 Z_6=\underline{C}_6-\frac{1}{2} \underline{C}_3\w H_3,
\end{equation}
and
\be T_5=\frac{1}{(2\pi)^5l_p^6},
\ee
is the tension of the M$5$-brane.
In the above action,
\begin{eqnarray}
\tilde H^{mn}=(\ast H)^{mnp}v_p,\\
H^{mn}=H^{mnp}v_p.
\end{eqnarray}
$H_{mnp}$ is the 3-form field strength in the worldvolume of the
M5-brane:
\begin{equation}
H_3=dA_2-\underline{C}_3,
\end{equation}
 and $v_p$ is defined by introducing an auxiliary field $b$:
\begin{equation}
 v_p=\frac{\p_p b}{\sqrt{g^{mn}\p_mb\p_nb}}.
\end{equation}
This auxiliary scalar field $b$ can be an arbitrary scalar with
nonzero gradient. We have made the choice that  the gradient of
$b$ is spacelike. The equation of motion of
  the auxiliary field $b$ is not independent. It can be obtained as
a consequence
  of the equation of motion of the 2-form gauge potential, which
  takes the following form after appropriate gauge fixing:
   \be\label{sdual}
   H_{mn}={\cal V}_{mn},
   \ee
   where
   \be
   {\cal V}_{mn}=-\frac{2}{\sqrt{-g}}\frac{\d\sqrt{-\det(g_{mn}+i{\tilde
   H}_{mn})}}{\d {\tilde H}^{mn}}.
   \ee
The relation (\ref{sdual}) can be understood as a generalized non-linear self-dual
condition.

 The ansatz of our M5-brane solution is the following:
as mentioned before, we take the $\Sigma_3$ part of the
worldvolume to be the same as the worldvolume of the above
membrane solution. The coordinates of this part are still chosen
to be $\xi_0, \xi_1, \xi_2$. As to the $\tilde S^3$ part, we
choose the worldvolume coordinates to be
\begin{equation}
\xi_3=\z_2,\hs{3ex} \xi_4=\z_3,\hs{3ex} \xi_5=\z_4,
\end{equation}
and we let $\z_1$ to be fixed at $\z^0$. We also make the
following ansatz for $d A_2$:
\begin{equation}
dA_2={R^3\over 8}\ta \sin^2\z_2\sin\z_3 d\z_2\w d\z_3 \w d\z_4,
\end{equation}
here $\ta$ is a constant. We choose the background three form
gauge potential to be
\begin{equation}
C_3={R^3\over 8}(3\cos\z_1-\cos^3\z_1) \sin^2\z_2\sin\z_3 d\z_2\w
d\z_3 \w d\z_4.
\end{equation}
so
\begin{equation} \underline{C}_3={R^3\over 8}(3\cos\z^0-\cos^3\z^0)
\sin^2\z_2\sin\z_3 d\z_2\w d\z_3 \w d\z_4.
\end{equation}

From now on, we will define $d(\z^0)$ as
\begin{equation}
d(\z^0)\equiv 3\cos\z^0-\cos^3\z^0,
\end{equation}
then
\begin{equation}
H_3={R^3\over 8}(\ta-d(\z^0)) \sin^2\z_2\sin\z_3 d\z_2\w d\z_3 \w
d\z_4.
\end{equation}
The hodge dual of $H_3$ is
\begin{equation}
\ast
H=\frac{\sqrt{-\mbox{det}g_{\Sigma_3}}(\ta-d(\z^0))}{\sin^3\z^0}d\xi_0
\w d\xi_1 \w d\xi_2.
\end{equation}
We choose the auxiliary scalar field $b$ to be $\xi_2$, then the
only nonzero component of $v^p$ is
$v^{\xi_2}=\sqrt{g^{\xi_2\xi_2}}$. So the only nonzero independent
compoent of $\tilde H_{mn}$ is
\begin{equation}
\tilde
H_{\xi_0\xi_1}=\sqrt{-\mbox{det}g_{\Sigma_3}g^{\xi_2\xi_2}}\frac{\ta-d(\z^0)}{\sin^3\z^0}.
\end{equation}
Then the first term of the PST action is: \bea T_5\int
d^6\xi\sqrt{-\mbox{det}(g+i\tilde H)}&=&\frac{T_5R^3}{8}\int
d^6\xi
\sqrt{-\mbox{det}g_{\Sigma_3}}\nn\\
&\times&\sin^2\z_2\sin\z_3\sqrt{\sin^6\z_0+(\ta-d(\z^0))^2}. \eea
It is easy to see that the second and the third terms of the PST
action vanish for our ansatz. So the PST action for our ansatz is:
\be S_{PST}=\frac{T_5R^3}{8}\int d^6\xi
\sqrt{-\mbox{det}g_{\Sigma_3}}\sin^2\z_2\sin\z_3\sqrt{\sin^6\z_0+(a-d(\z^0))^2}.
\ee We need to find the value of $\z^0$ such that the action take
the minimal value. Define \be x\equiv\cos\z^0, \ee and \be
f=\sin^6\z^0+(a-d(\z^0))^2=(1-x^2)^3+(\ta-3x+x^3)^2. \ee From
${df\over dx}=0$, we get $x=\ta/2$.\footnote{The other two
solutions of $df/dx=0$, $x=\pm 1$, will give us the M$5$-brane
solutions with shrinking $\tilde S^3$. We will not consider these
solutions here. } Then \be H=-{R^3\over
8}\cos\z^0\sin^2\z^0\sin^2\z_2\sin\z_3 d\z_2\w d\z_3\w d\z_4.\ee
Now the action of M5-brane equal to the volume of $\Sigma_3$ times
a constant. Then $\Sigma_3$ should be a $3$-dimensional
submanifold with minimal volume. It is guaranteed by the fact that
$\Sigma_3$ is the worldvolume of a M$2$-brane whose configuration
is the solution of the membrane equations of motion. So our ansatz
does satisfy the M5-brane equations of motion when $\ta$ and
$\z^0$ satisfy \be \cos\z^0=\frac{\ta}{2}. \ee

Using eqs.~(\ref{eqH3}) and (\ref{az}), one can find that the
$\tilde S^3$ part of $H_3$ of the M$5$-brane solution in the
previous subsection is the same as the obtained $H_3$ in this
section. This show that that M$5$-brane solution is a special case
of the universal result of this section.\footnote{In the PST
formalism, the self-dual condition is eq.~(\ref{sdual}) which is
from the equations of motion. We need not to ask $H_3$ to be
constructed from a self-dual $3$-form $h_3$ on the worldvolume of
the M$5$-brane as what we did in the previous subsections. This is
the reason why the $H_3$ in this subsection does not have a part
along the directions in $M_7$. } Another special case was studied
in \cite{Chen}.

As a nontrivial check of this universal result, we have studied
the following ansatz for membrane in $Sch.$-$AdS_7$ space:
 \be
 \xi_0=t,~~ \xi_1=x, ~~\xi_2=y,~~r=g(y). \ee
This ansatz is just the $Sch.$-$AdS_7$ part of the M$5$-brane
ansatz eq.~(\ref{m5brane2}) in the previous section. The membrane
equations of motion for this ansatz give the same constraint on
$g$ as the one obtained from the M$5$-brane equations,
eq.~(\ref{eg}).

Using this universal result, one can easily obtained the M5-brane
configurations corresponding to two parallel straight
Wilson-Polyakov surfaces in the same higher anti-symmetric
representation from the M2-brane configurations discussed in
subsection \ref{potential}.

\section{Conclusion and discussions}

In this paper, we investigated the thermodynamical behaviors of
six-dimension $(2, 0)$ field theory by studying the
Wilson-Polyakov surface operators in this theory. We proposed that
these operators should be described by M-theory branes.  When
these operators are in low dimensional representations, M2-brane
configuration is a good description. While if these operators are
in higher dimensional representation, we suggested that a better
description should be in terms of M5-branes. We used our membrane
description to study the interaction potential between two strings
and found that when the distance between them is small, the
potential's behaviors are asymptotically similar to
zero-temperature results \cite{Mal98}, while if the distance is
large enough the interaction will be screened by the finite
temperature effects. Qualitatively this result is similar to the
potential between two quarks in the four dimensional
SYM\cite{Polyakovloop}.

Although the M5-brane solution dual to straight Wilson-Polyakov
surfaces in anti-symmetric representations are not very hard to
find. Searching for the M5-brane solution dual to the
Wilson-Polyakov surfaces in symmetric representation leads to a
quite complicated differential equation. We discussed the existence
of the solution and showed that when the temperature is small
enough, the M5-brane solution should exist.

Inspired by our study of M5-branes dual to Wilson-(Polyakov)
surfaces, we proved a universal result on M5-brane solution in a
quite generic background $M_7\times S^4$ with four-form flux.
Given any membrane solution in this background with worldvolume
$\Sigma_3$ completely embedded in $M_7$, we get an M5-brane
solution with topology $\Sigma_3\times \tilde S^3$ with $\tilde
S^3$ being in $S^4$. We hope that this universal result is useful
to study the dynamics of M-theory branes in generic background
noticing that supersymmetries play no roles at all here. We hope
that this results will also be useful in probing some other
six-dimensional theory which has a gravity dual.

Quite less is known about the six-dimensional superconformal field
theory. This theory at finite temperature theory is even less studied. As
mentioned in the introduction, by compacting on a two-torus, the
six dimensional theory will reduced to four dimensional ${\cal
N}=4$ super Yang-Mills theory at low energy. If we wrapping the
Wilson-Polyakov surface on a suitable circle of this two-torus, we
expect to get the Wilson-Polyakov loop. Hope that this relation
will tell us more about the thermodynamics of this six-dimensional
theory in the future.

It would be interesting to study more thoroughly the properties of
the six-dimensional superconformal field theory at the finite
temperature. The theory could be in a phase of perfect fluid, just
like the quark-gluon plasma phase of ${\cal N}=4$ super-Yang-Mills
theory at finite temperature. Then one can use the AdS gravity to
study the physics in this phase\cite{ChenWu3}.

\section*{Acknowledgments}

The work was partially supported by NSFC Grant No. 10535060,
10775002 and NKBRPC (No. 2006CB805905). We would like to thank Kumar
S. Narain for suggesting this problem to us and quite useful
discussions. JW also thanks Antonio Ambrosetti, Matteo Bertolini,
Loriano Bonora, Edi Gava, Xiaoli Han, Jiayu Li, Bin Liu and Hong-Jun
Yu for helpful discussions. BC would like to thank OCU and KEK for
their hospitality, where part of this project was carried on.

\section{Appendix: Various connections}

  In this appendix, we list various connections appeared in our
  calculation. For the induced metric (\ref{indmetric}), its Christoffel symbol
  has nonvanishing independent components:
  \bea
  \G^t_{yt}&=&\frac{1}{2}(\frac{f^{\pr}}{f}-\frac{2}{y}), \nn \\
  \G^x_{yx}&=&-\frac{1}{y}, \nn \\
  \G^y_{tt}&=&\frac{f}{2(1+fg^{\pr 2})}(-\frac{2f}{y}+f^\pr), \nn\\
  \G^y_{xx}&=&\frac{f}{(1+fg^{\pr 2})y},\nn\\
  \G^y_{yy}&=&-\frac{1}{y}+\frac{1}{1+fg^{\pr 2}}(-\frac{f^\pr}{2f}+f {g^\pr} {g^{\pr \pr}}), \nn \\
  \G^y_{\a\a}&=&\frac{fg^2}{1+fg^{\pr 2}}(\frac{1}{y}-\frac{g^{\pr}}{g}),\nn\\
  \G^y_{\b\b}&=&\G^y_{\a\a}\sin^2\a,\nn\\
  \G^y_{\g\g}&=&\G^y_{\a\a}\sin^2\a\sin^2\b, \nn\\
   \G^\a_{y\a}&=&\G^\b_{y\b}=\G^\g_{y\g}=(-\frac{1}{y}+\frac{g^{\pr}}{g}),\nn\\
  \G^\a_{\b\b}&=&-\sin\a\cos\a, \nn\\
   \G^\a_{\g\g}&=&-\sin^2\b\sin\a\cos\a, \nn\\
  \G^\b_{\a\b}&=&\G^{\g}_{\a\g}=\frac{\cos\a}{\sin\a},\nn\\
  \G^\b_{\g\g}&=&-\sin\b\cos\b, \nn\\
   \G^\g_{\b\g}&=&\frac{\cos \b}{\sin\b}.
  \eea

  For the $Sch.$-$AdS_7$ spacetime, its nonvanishing independent components of spin connection
   are
  \bea
  \o^{\underline 2}_{{\underline 0}{\underline 0}}&=&\frac{y^2}{R}\p_y(\frac{\sqrt{f}}{y}),
  ~~~\o^{\underline 2}_{{\underline i}{\underline
  i}}=\frac{\sqrt{f}}{R}, \hspace{3ex}\mbox{for $i\neq 0,2$},\nn\\
  \o^{\underline 3}_{\underline ii}&=&-\frac{y}{Rr},\hspace{3ex}\mbox{for $i=4,5,6$},\nn\\
  \o^{\underline 4}_{{\underline i}{\underline
  i}}&=&-\frac{y\cos\a}{Rr\sin\a},\hspace{3ex}\mbox{for $i=5,6$},\nn\\
  \o^{\underline 5}_{{\underline 6}{\underline
  6}}&=&-\frac{y\cos\b}{Rr\sin\a\sin\b}.
  \eea
And its nonvanishing independent components of Christoffel symbol
are \bea
\G^{\underline y}_{{\underline y}{\underline y}}&=&-\frac{1}{y}-\frac{f^\pr}{2f},\nn\\
\G^{\underline y}_{{\underline t}{\underline t}}&=&\frac{1}{2}f
f^\pr -\frac{f^2}{y},\nn\\
\G^{\underline y}_{{\underline x_i}{\underline
x_i}}&=&\frac{f}{y}, \hspace{3ex}\mbox{for $i=1, \cdots, 5$},\nn\\
\G^{\underline t}_{{\underline y}{\underline t}}&=&-\frac{1}{y}+\frac{f^\pr}{2f},\nn\\
\G^{\underline x_i}_{{\underline y}{\underline
x_i}}&=&-\frac{1}{y}. \eea

\end{document}